# Non-homogeneous structure of complex concentrated alloys: Effect of intrinsic strain

Václav Paidar[1], Pavel Lejček[1], Andrea Školáková[1,*]

[1] FZU – Institute of Physics of the Czech Academy of Sciences, Na Slovance 1999/2, Prague 8 182 00, Czech Republic

*Corresponding author: skolakova@fzu.cz

## Abstract

Even if the atoms of a multicomponent alloy occupy a common lattice, their distribution is not homogeneous, and regions with different compositions can be detected. Three representative examples will be discussed: a Cantor-type system containing transition-metal elements (Cr, Mn, Fe, Ni, and Co), a refractory high-entropy alloy (Ti, Zr, Nb, Ta, and Mo), and a multicomponent system combining transition and refractory metals (Cu, Ni, Ti, Zr, and Hf). Using a combination of theoretical analysis and experimental observations, we demonstrate that the formation of locally segregated regions can lead to a reduction in the overall energy of the system. This stabilization arises from the compensation of tensile and compressive strain fields associated with atoms of different sizes, highlighting the key role of local chemical and structural heterogeneity in determining the thermodynamic stability of multicomponent alloys.

In addition to single phase multicomponent alloys, there is a large number of multicomponent alloys that are not homogeneous. These compositional variations are not only of fundamental interest but also have important implications for the stability and mechanical behaviour of multicomponent alloys. Understanding these effects motivates a detailed investigation of local chemical heterogeneities in multicomponent alloys. Let us consider three examples: a quinary system of the Cantor type (equimolar CrMnFeCoNi), which exhibits high strength and ductility [1], a refractory system (equimolar TiZrNbTaMo), with superior mechanical properties surpassing those of conventional superalloys [2], and Cu–Ni–Ti–Zr–Hf-based alloys, showing shape-memory behaviour [3]. In the latter case, an alloy of composition of Cu15Ni35Ti25Zr12.5Hf12.5 (at.%) is considered. These three compositions represent some of the most typical examples of multicomponent alloys, now more commonly referred to as complex concentrated alloys (CCA). They are particularly suitable for our analysis, as they include a representative system based solely on transition metals, a system based on refractory elements, and a system combining both types of elements.

A five-component transition-metal CCA (20.5 Cr, 19.9 Mn, 18.9 Fe, 20.1 Co, 20.6 Ni, at.%) forms a single fcc solid solution which solidifies dendritically. However, after annealing the original equi-atomic composition is losing up its character of homogeneous high-entropy multicomponent system at grain boundaries [4], since at certain regions only nickel and manganese atoms are strongly prevailing [5]. Those two elements have high enthalpy of mixing, -8.2 kJ/mol [6]. Slightly lower values of mixing enthalpy are for Ni–Cr (-6.7 kJ/mol) and Mn–Co (-5.2 kJ/mol) followed by Co–Cr (-4.5 kJ/mol) [6]. Those binary systems do not form solid solutions.

For the evaluation of energy variations, we need to know the atomic volumes, $V$, and compressibility (bulk moduli), $K$. For simple comparison, let us use the lattice parameters of the bcc lattice. For the fcc metals we determine hypothetical bcc lattice parameters from their atomic volumes, $(a_{bcc})^3 = (a_{fcc})^3/2$. Similarly, for the hcp metals we use $(a_{bcc})^3 = (a_{hcp})^2 c_{hcp}\sqrt{2/3}$. All basic parameters are summarized in Table 1.

**Table 1.** Lattice parameters and bulk moduli of three considered systems [7-9].

| | Cr | Mn | Fe | Co | Ni |
|---|---|---|---|---|---|
| bcc lattice parameter (pm) | 288 | 298 | 286 | 281 | 280 |
| $K$(GPa) | 160 | 120 | 170 | 180 | 180 |

| | Ti | Zr | Nb | Ta | Mo |
|---|---|---|---|---|---|
| bcc lattice parameter (pm) | 328 | 360 | 330 | 330 | 315 |
| $K$GPa) | 110 | 91 | 170 | 200 | 230 |

| | Ti | Zr | Hf | Ni | Cu |
|---|---|---|---|---|---|
| bcc lattice parameter (pm) | 328 | 360 | 355 | 280 | 287 |
| $K$(GPa) | 110 | 91 | 110 | 180 | 140 |

Let us suppose that in the initial multicomponent alloy, all elements possess the average atomic volume, it means that the distances between the smaller atoms are extended and those between the larger atoms are contracted. The energies of individual atoms can be evaluated as

$$E_{def} = VK(\Delta V/V). \quad (1)$$

The compositions and corresponding deformation energies are summarized in Table 2. The positive signs indicate contractions while the negative signs extensions. To be consistent with common thermodynamic data, the energies are not presented per atoms but in kJ/mol.

**Table 2.** Initial compositions and deformation energies of individual atoms.

| | Cr | Mn | Fe | Co | Ni |
|---|---|---|---|---|---|
| initial composition | 21 | 20 | 19 | 20 | 21 |
| $E_{def}$ (kJ/mol) | 20 | 108 | -4 | -69 | -108 |

| | Ti | Zr | Nb | Ta | Mo |
|---|---|---|---|---|---|
| initial composition | 20 | 20 | 20 | 20 | 20 |
| $E_{def}$ (kJ/mol) | -48 | 267 | -43 | -51 | -383 |

| | Ti | Zr | Hf | Ni | Cu |
|---|---|---|---|---|---|
| initial composition | 25 | 12.5 | 12.5 | 35 | 15 |
| $E_{def}$ (kJ/mol) | 26 | 334 | 341 | -677 | -456 |

In the binary phase diagram Mn–Ni, there is a B2 phase in the composition interval 45–51 at.% Ni for the temperatures below solidus. It indicates that the contraction deformation of manganese atoms and the extension deformation of nickel atoms can be mutually compensated and, in that way, the average deformation energy of 62 kJ/mol per atom can be reduced to 19 kJ/mol. It results in the formation of Mn–Ni regions surrounded by Cr–Fe–Co regions. This behaviour has been observed on grain boundaries in the high-entropy alloys in [5].

As a second example of non-homogeneous multicomponent system let us discuss the refractory system Ti–Zr–Nb–Ta–Mo [2]. Non-homogeneous character of high-entropy alloys has been pointed out already by Miracle and Senkov [10, 11].

Starting with the atomic volumes, we can define the lattice parameters of hypothetical bcc lattices also for the hexagonal metals (see Table 1). Let us distinguish two groups of elements, the first one corresponding to dendritic parts of the structure, namely, Mo and Ta, and the second group corresponding to inter-dendritic regions, namely, Ti, Nb and Zr. In the non-annealed samples, we estimate only the average compositions, nevertheless, in the annealed samples the dendritic (DE) and inter-dendritic (ID) regions are considered separately. Moreover, a transition region between DE and ID regions can be introduced as well.

In the third case of Cu-Ni-Ti-Zr-Hf alloy, the regions were classified, based on the results of the EDS (Energy Dispersive Spectroscopy) line scan, into dark region containing Ti and Zr, and bright regions containing Ni, Cu, and Hf. The evaluated compositions, derived from variations in the BSE contrast and line-scan EDS analysis, are summarized in Table 3. The data for the Ti-Zr-Nb-Ta-Mo alloy are taken from our previous work [2], while the corresponding results for the Cu-Ni-Ti-Zr-Hf alloy are shown in Fig. 1. These values reflect relative elemental differences rather than exact chemical concentrations.

**Table 3.** Relative compositions (in at.%).

| | Ti | Zr | Nb | Ta | Mo |
|---|---|---|---|---|---|
| non-annealed | 12 | 24 | 30 | 8 | 26 |
| dendrites | 14 | 17 | 29 | 9 | 31 |
| transition | 10 | 36 | 33 | 4 | 17 |
| inter-dendrites | 9 | 44 | 34 | 2 | 11 |
| | Ti | Zr | Hf | Ni | Cu |
| initial | 25 | 12.5 | 12.5 | 35 | 15 |
| bright | 24 | 14 | 30 | 21 | 11 |
| dark | 39 | 24 | 15 | 13 | 9 |

Let us discuss first the case of Ti-Zr-Nb-Ta-Mo. During annealing the concentration of molybdenum atoms increases in DE zones. Similarly, also the compositions of tantalum and titanium are slightly higher in the DE zones and simultaneously smaller in the ID zones. On the other hand, the concentrations of zirconium and niobium are higher in the ID zones. Crucial changes are those of molybdenum and zirconium, while the molybdenum is concentrated in the DE zones, the zirconium is depleted from them.

When the DE zones are principally composed of molybdenum and tantalum, their bcc lattice parameter can be evaluated as 319 pm and the related hypothetical bcc lattice parameter of the ID zones, composed principally of zirconium, niobium and titanium, is 345 pm. These numbers are close to the measured values of two bcc structures, namely, 327 and 349 pm [2]. For the third multicomponent system we have found the same bcc lattice parameter 322 pm, namely, for dark region enriched in Ti and Zr, as well as for the bright region containing Cu, Ni, and Hf. Our estimates of the hypothetical bcc lattice parameter according to the compositions of Table 3 are 330 pm for dark regions and 326 pm for bright regions, respectively.

Similarly, as in the case of transition-metal systems, let us suppose that all the atoms are initially deformed in such a way that all of them possess the same atomic volume. The average lattice parameter of hypothetical bcc structure for the refractory system is then equal to 333 pm. Hence only the zirconium atoms are larger than the average. In the same way as for transition-metal alloys, the energy necessary for the unifying atomic deformations can be estimated when the relative deformation is multiplied by the bulk modulus and atomic volume.

The large hypothetical deformation energies for molybdenum and zirconium, summarized in Table 2, are a crucial reason for the formation of the bcc structures with two different lattice parameters. In the Cu-Ni-Ti-Zr-Hf system, similar effects are also relevant in the formation of the monoclinic phase. Although both dendritic regions exhibit the same calculated bcc lattice parameters (322 pm), the phase formation is not determined solely by lattice parameter matching. Instead, thermodynamic and electronic likely destabilize the bcc solid solution, promoting the formation of monoclinic phase.

The mixing enthalpies are positive for the pairs of elements that do not mix, Ti–Nb (2.0 kJ/mol), Ti–Ta (1.4 kJ/mol), Zr–Nb (3.9 kJ/mol), Zr–Ta (2.7 kJ/mol) [6]. The solubility of titanium or zirconium in both niobium or tantalum is limited [8, 12]. In between the concentrations, where a bcc element is soluble in the hexagonal lattice and where a hexagonal element is soluble in the bcc lattice, there is a miscibility gap where the alloy is separated into regions with different concentrations.

The binary phase diagrams of hexagonal and transition-metal elements are characterized by miscibility gaps, that can be described by their width in at.% and the temperature difference between the gap maximum and the monotectoid temperature. The efficiency for the element separation can be characterized by the product of the gap width and the gap hight. This quantity is large for the pairs of zirconium and tantalum and for titanium and niobium.

The deformation energies for the individual atoms in different regions (DE, ID, dark, bright) are presented in Table 4. Their magnitudes are comparable (in summary 187 for dendrites and 180 kJ/mol for inter-dendrites) for the refractory alloys, Ti–Zr–Nb–Ta-Mo. The large deformation energies of molybdenum and zirconium are located in two different regions, namely, in dendrites and inter-dendrites.

The behaviour of the Cu-Ni-Ti-Zr-Hf system is similar. The deformation energies in summary are partly different for the dark and bright regions (271 and 348 kJ/mol). But the large deformation energies of the pairs nickel/zirconium and nickel/hafnium are mutually partially compensated. Moreover, the elements that well mix (titanium/zirconium) are separated from the other elements.

**Table 4.** Deformation energies, $E_{def}$, (kJ/mol) in different regions.

| | Ti | Zr | Nb | Ta | Mo |
|---|---|---|---|---|---|
| Dendrite | -7 | 46 | -13 | -5 | -117 |
| inter-dendrite | -4 | 118 | -15 | -1 | -42 |
| | Ti | Zr | Hf | Ni | Cu |
| Dark | 10 | 80 | 51 | -88 | -41 |
| Bright | 6 | 47 | 102 | -142 | -50 |

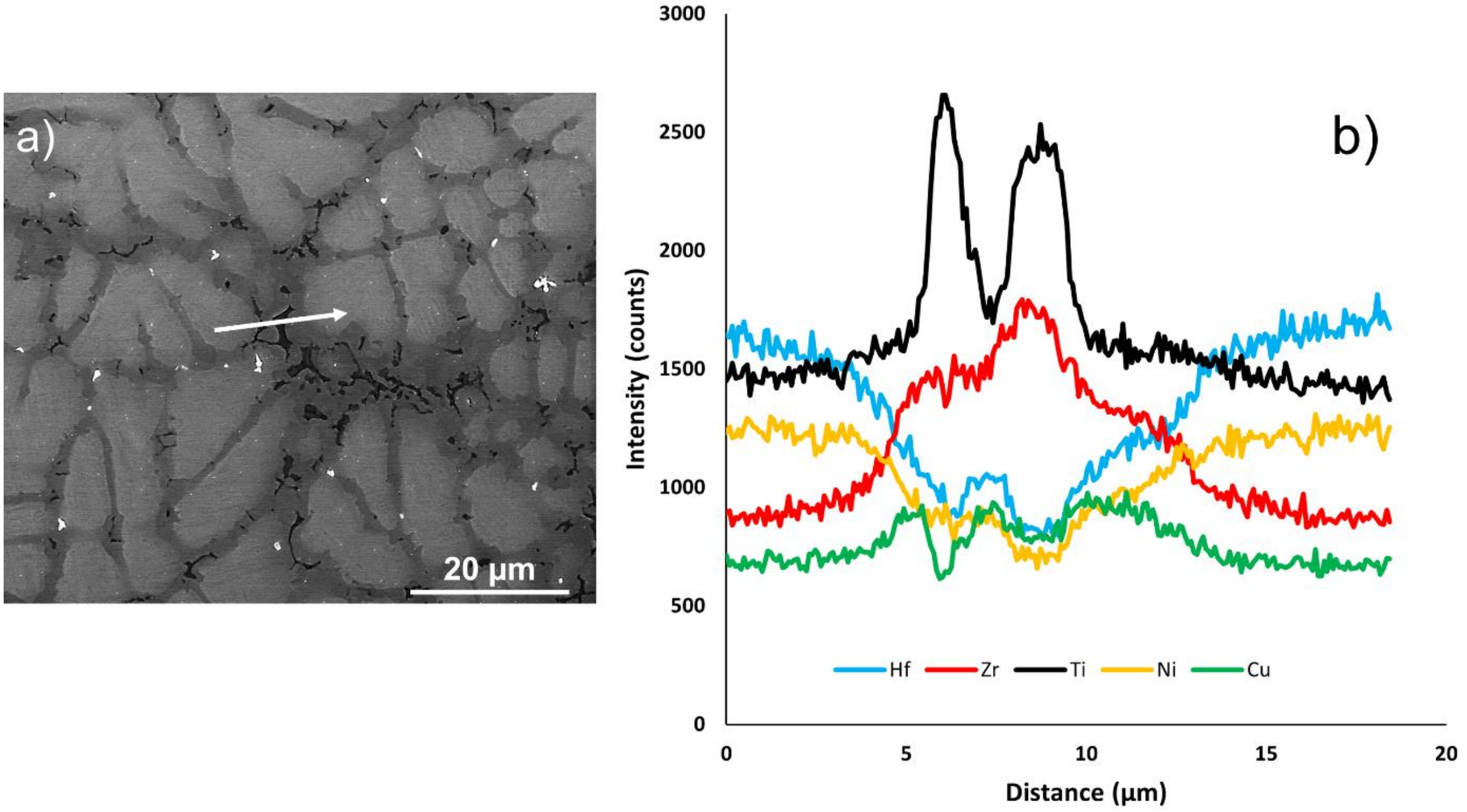

**Fig. 1:** a) Backscattered electron (BSE) scanning electron microscopy (SEM) image of the Cu15Ni35Ti25Zr12.5Hf12.5 alloy with an arrow indicating the direction of the line scan analysis; b) EDS line scan showing chemical composition variation.

In the binary phase diagram, Ti and Zr are fully miscible. On the other hand, Ti or Zr with Ni form several intermetallic compounds, similarly, also Ti or Zr with Cu form intermetallic compounds. Hf with Ni or Cu belong to the binary systems that do not mix as well.

In summary, three examples of complex concentrated alloys have been discussed; their non-homogeneous structure can occur for different reasons. In all cases, the driving force for structure separation is related to large differences of atomic volumes of individual atomic components associated with large deformation energies.

**Declaration of Generative AI Use**

During the preparation of this work the authors did not use any tool of AI.

**CRediT authorship contribution statement**

**Václav Paidar:** Conceptualization, Writing – review & editing, Writing – original draft, Investigation, Formal analysis. **Pavel Lejček:** Writing – review & editing, Formal analysis. **Andrea Školáková:** Writing – review & editing, Writing – original draft, Investigation, Formal analysis, Funding acquisition.

**Declaration of Interest Statement**

The authors declare that they have no known competing financial interests or personal relationships that could have appeared to influence the work reported in this paper.

**Acknowledgment**

The authors acknowledge financial support from the Czech Science Foundation (Grant No. 26-21374S). The authors acknowledge the assistance provided by the Ferroic Multifunctionalities project, supported by the Ministry of Education, Youth, and Sports of the Czech Republic. Project No. CZ.02.01.01/00/22_008/0004591, co-funded by the European Union. CzechNanoLab project LM2023051 funded by MEYS CR is gratefully acknowledged for the financial support of the measurements/sample fabrication at LNSM Research Infrastructure.

**Data availability**

The used data are accessible via the Zenodo repository: https://doi.org/10.5281/zenodo.20040700.